\documentclass[preprint,nofootinbib,nobibnotes,groupaddress,preprintnumbers,aps,prd]{revtex4}
\usepackage{bm,epsfig,mathrsfs,amsmath,amssymb,graphicx}
\usepackage{epsfig,amsmath,graphicx,amssymb}
\usepackage{graphicx}
\usepackage{dcolumn}
\usepackage{bm}

\begin{document}

\title{ Coefficient of performance for  a low-dissipation Carnot-like refrigerator with  nonadiabatic dissipation}

\author{Yong Hu$^{1}$}
\author{Feifei Wu$^{1}$}
\author{Yongli Ma$^{2}$}
\author{Jizhou He$^{1}$}
\author{Jianhui Wang$^{1,2}$}  \email{wangjianhui@ncu.edu.cn}
\author{A. Calvo Hern\'{a}ndez$^{3}$}
\author{J. M. M. Roco$^{3}$}
\affiliation{ $^1\,$ Department of
Physics, Nanchang University, Nanchang 330031, China \\
$^2\,$ State Key Laboratory of Surface Physics and Department of
Physics, Shanghai 200433,  China} \affiliation{$^3\,$ Departamento
de F\'{i}sica Aplicada and Instituto Universitario de F\'{i}sica y
Matem\'{a}ticas (IUFFYM), Universidad de Salamanca, 37008 Salamanca,
Spain }

\begin{abstract}
 We study the
coefficient of performance (COP) and its bounds of the Canot-like
refrigerator working between two heat reservoirs at constant
temperatures $T_h$ and $T_c$, under two optimization criteria
 $\chi$ and $\Omega$. In view of the fact that an ``adiabatic'' process
takes finite time and is nonisentropic, the nonadiabatic dissipation
and the finite time required for the ``adiabatic'' processes are
taken into account. For given optimization criteria, we find that
the lower and upper bounds of the COP are the same as the
corresponding ones obtained from the previous idealized models where
any adiabatic process undergoes instantaneously with constant
entropy. When the dissipations of two ``isothermal'' and two
``adiabatic'' processes are symmetric, respectively, our theoretical
predictions match the observed COP's of real refrigerators more
closely than the ones derived in the previous models, providing a
strong argument in favor of our approach.


PACS number(s): 05.70.Ln

\end{abstract}

\maketitle
\date{\today}
%%%%%%%%%%%%%%%%%%%%%%%%%%%%%%%%%%%%%%%%%%%%%%%%%

\section {introduction}
The issue of thermodynamic optimization on cyclic  converters has
attracted much attention because of the sustainable development in
relation to any energy converter operation.  Along this issue, a
number of various performance regimes \cite{Wu04, Ber00, Dur04} have
been considered within different figures of merit to disclose
possible universal and unified features, with special emphasis on
the possible consistency between theoretical predictions and
experimental data. If heat engines, or  refrigerators and heat
pumps, works between two heat reservoirs at constant temperatures
$T_h$ and $T_c$, practically they  operate far from the ideal
maximum Carnot efficiency ($\eta_{\mathrm{max}}=\eta_C=1-T_c/T_h$),
or the maximum Carnot coefficient of performance
(COP)[$\varepsilon_{\mathrm{max}}=\varepsilon_C=T_c/(T_h-T_c)$],
which requires an infinite time to complete a cycle. By contrast,
the maximum output for heat engines, or  the maximum cooling rate
for refrigerators and maximum heating rate for heat pumps, can be
achieved within finite cycle time. In most studies of the
Carnot-like heat engine models, the power output  as a target
function is always maximized to find valuable and simple expressions
of the optimized efficiency \cite{Bro05, Izu08, Esp09, Tu08, Zctu12,
Esp10, Guo13, Huang13, Sei08, Sei11}. Without assuming any specific
heat transfer law or the linear-response regime, Esposito \emph{et
al.} \cite{Esp10} proposed the low-dissipation assumption that the
irreversible entropy production in a heat-exchange process is
inversely proportional to the time spent on the corresponding
process, and they re-derived the paradigmatic Curzon-Ahlborn value
\cite{Cur75} $\eta_{CA} =1-\sqrt{1-\eta_C}$ in the  limit of symmetric dissipation. %without assuming any specific heat transfer
%law or the linear-response regime.%model for low-dissipation Carnot-like engines, by
%considering a low-dissipation Carnot
%heat engine model   %which under endoreversible
%assumptions (i.e., all considered irreversibilities coming from the
%couplings between the working system and the external heat
%reservoirs through linear heat transfer laws) recover the
%paradigmatic Curzon-Ahlborn value [16] ¦ÇmaxP =1. ¡Ì ¦Ó ¡Ô¦ÇCA using
%¦Ó ¡ÔTc/Th, where Tc and Th denote the temperature of the cold and
%hot heat As a consequence, the optimization of heat devices based on
In addition to the power output, the per-unit-time efficiency, a
compromise between the efficiency and the speed of the whole
heat-engine cycle, was considered as another criterion \cite{Ma85}
of optimization.

It is more difficult to adopt a suitable optimization criterion and
determine its corresponding COP for refrigerators, in comparison
with dealing with issue of the efficiency at maximum power for heat
engines. Various optimization criteria \cite{Yan90, Vel97, All10,
Tom13, Chen95, Tom12R, Tu12} have been proposed in optimum analysis
of a classical or quantum refrigeration cycle. Chen and Yan
\cite{Yan90} introduced the function $\chi=\varepsilon Q_c/\tau$,
with $Q_c$ the heat transported from the cold reservoir and $\tau$
the cycle time, as a target function within
finite-time-thermodynamics context. Velasco \emph{et al.}
\cite{Vel97} adopted the per-unit-time COP as a target function
while Allahverdyan \emph{et al.} \cite{All10} introduced
$\varepsilon Q_c$ to be the target function.  C.de Tom\'{a}s
\emph{et al.} \cite{Tom12R} proved the COP at maximum $\chi$ for
symmetric low-dissipation refrigerators to be
$\varepsilon_{CA}=\sqrt{\varepsilon_C+1}-1$, where
$\varepsilon_C=T_c/(T_h-T_c)$ is the Carnot COP.  Based on the
$\chi$ figure of merit, Wang \emph{et al.} \cite{Tu12} obtained the
lower and upper bounds of the COP and showed that these bounds can
be achieved in extremely asymmetric dissipation limits. Very
recently, C. de Tom\'{a}s \emph{et al.} \cite{Tom13} studied the
low-dissipation heat devices and obtained the bounds of COP under
general and symmetric conditions, by applying the unified $\Omega$
optimization criterion, which was first proposed in \cite{Her01} to
consider a compromise between energy benefits and losses for a
specific job. This criterion  has been applied to the performance
optimization on a wide variety of energy converters \cite{San03,
Jim08, San10}.

Most of the previous studies about the performance in finite time of
heat devices did not take into account nonadiabatic dissipation for
the cyclic converter by assuming that the adiabatic steps run
instantaneously with constant entropy, though the importance of
nonadiabatic dissipation in an adiabatic process was suggested by
Novikov \cite{Nov58}. The influence on the performance of a
classical or quantum heat engine, induced by internally dissipative
dissipation  (such as inner friction and internal dynamics,
\emph{etc.}),
 has been discussed in several papers \cite{Rui13, Fel00, pre86,
 pre85,
 Chen94, Gor91, Ber12, Ape12}. To the best of our knowledge, so far  little
attention has been paid to the effects of nonadiabatic dissipation
on the performance characteristics of the refrigerators proceeding
with finite time. It is therefore of significance to consider a more
generalized refrigerator model by involving the nonadiabatic
dissipation and the time spent on an adiabatic process.

 In the present paper, we consider a low-dissipation Carnot-like
 refrigeration cycle
 of two irreversible isothermal and two irreversible adiabatic processes, and
analyze its COP at the $\chi$ and $\Omega$ figures of merit,
respectively. We show that the inclusion of adiabatic dissipation
does not lead to any change in the bounds of the COP at a given
figure of merit, as expected. When the dissipations of the two
isothermal and two adiabatic processes are symmetric, we find that,
 our results agree well with the data of
the real refrigerators, thereby indicating that inclusion of
nonadiabatic dissipation is essential. Throughout the paper, we use
the word ``isothermal"  to mean that the working substance is
coupled to a reservoir with constant temperature, while we adopt the
word ``adiabatic" to indicate merely that no heat exchanges between
the working substance and its surroundings.
\section{model}
\begin{figure}[h]
\includegraphics[width=300pt]{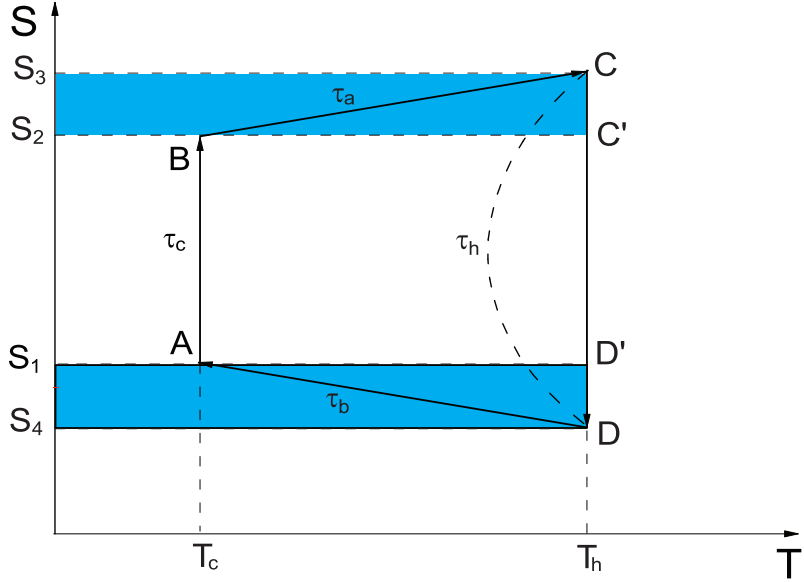}
\caption{(Color online) Schematic diagram of an irreversible
Carnot-like refrigeration cycle in the plane of the temperature $T$
and entropy $S$. The values of the entropy $S$ at the four special
instants are indicated by $S_i$ ($i=1, 3, 3, 4$). Here $\tau_{h,~c}$
are the times spent on the two isothermal process, while $\tau_{a,~
b}$ represent the times taken for the two adiabatic processes. The
blue area defined by the rectangles $C, C', S_2, S_3$ and $D, D',
S_1, S_4$ represents the total work done by the system to overcome
nonadiabatic dissipation  in the two adiabatic processes. }
\label{st}
\end{figure}

An irreversible Carnot-like refrigeration cycle $A\rightarrow
B\rightarrow C\rightarrow D\rightarrow A$ is drawn in the $(S, T)$
plane (see Fig. \ref{st} ). During two isothermal processes
$A\rightarrow B$ and $C\rightarrow D$, the working substance is in
contact with a cold and a hot heat bath at constant temperatures
$T_h$ and $T_c$, respectively.  In the adiabatic process
$B\rightarrow C$ ($D\rightarrow A$), the working substance is
decoupled from the cold (hot) reservoir, and the entropy changes
from $S_B$ to $S_C$ ($S_D$ to $S_A$). It can be seen from Fig.
\ref{st} that $S_1=S_A$, $S_2=S_B$, $S_3=S_C$ and $S_4=S_D$. % Let
%$S_i$ be the entropies at the instants $i$ with $i=A, B, C, D$.
For
the reversible cycle where $S_B=S_C$ and $S_A=S_D$, we recover the
Carnot efficient of performance
$\varepsilon{_C}=\frac{T_c}{T_h-T_c}$, which is generically
universal.

  Now we turn to discussion on the Carnot-like cycle under finite-time
  operation that moves the working substance away from the
  equilibrium. In the isothermal process the system may be out of
  equilibrium, but it must be in the equilibrium
   with the heat reservoir at the special  instants $i$ with $i=A, B,
C, D$, at which the thermodynamic quantities of the system can be
defined well.  Unlike in the ideal case where any adiabatic process
is isentropic, the adiabatic process is nonisentropic because of
nonadiabatic dissipation.  This dissipation develops additional heat
and thus yields  an
increase in the entropy during the so-called adiabatic process.  %This additional heat remains in a chamber or in  a trap
%along an adiabatic process until it is released to a heat reservoir
%with which the working systems couples during an isothermal process.
%As a consequence, heat productions due to friction in the adiabatic
%expansion $B\rightarrow C$ and  in the adiabatic compression
%$D\rightarrow A$ are released into the cold and hot reservoirs,
%respectively \cite{Ape12, Wu92}. This additional  heat is also
%represented in Fig. \ref{st} by the red triangular area for the
%branch $B\rightarrow C$ and by the blue triangular area for the
%process $D\rightarrow A$. Note that, the heat produced during the
%process $D\rightarrow A$ decreases the absorbed heat during the hot
%isothermal process, and the heat produced during the adiabat
%$B\rightarrow C$, as pure loss, is released to the cold reservoir.
The irreversible Carnot-like refrigerator that consists of the two
adiabatic and  two isothermal processes is operated as follows (more
details about the isothermal processes can be seen in \cite{Tu12}).

1. Isothermal expansion $A \to B$. The working substance is in
contact with the cold reservoir at temperature $T_c$ for a period
$\tau_c$. In this expansion the constraint imposed on the system is
loosened according to the external controlled parameter
$\lambda_c(\tau)$ during the time interval $0 <\tau < \tau_c$, where
$\tau$ is the cycle-time variable. A certain amount of heat $Q_c$ is
released to the cold reservoir and the variation of entropy can be
expressed as
\begin{equation}
\Delta S_c  = Q_c /
 T_c  + \Delta S_c^{ir}
\label{ts1},
\end{equation}
with $\Delta S_c^{ir}\geq0 $ being the irreversible entropy
production.

2. Adiabatic compression $B \to C$.  The entropy is increased due to
irreversible entropy production caused by the nonadiabatic
dissipation, while the constraint on the system is  enhanced
according to the external controlled parameter $\lambda_a(\tau)$
during the time interval $\tau_c < \tau < \tau_c+\tau_a$. The
irreversible entropy production arising from the nonadiabatic
dissipaton is denoted by
\begin{equation}
\Delta S_a^{ir}  = S_3 - S_2 \label{ps1}.
\end{equation}

3. Isothermal compression $C \to D$. The working substance is
coupled to a hot reservoir at constant temperature $T_h$ for time
$\tau_h$. The constraint on the system is further  enhanced with the
external controlled parameter $\lambda_h(\tau)$ during the time
interval $\tau_c+\tau_a < \tau <\tau_c+\tau_a+\tau_h$.  Let $Q_h $
be
  an amount of heat released to the hot reservoir, we have the entropy
  variation,
\begin{equation}
\Delta S_h  =   - Q_h/
 T_h + \Delta S_h^{ir}
\label{ts2},
\end{equation}
where $\Delta S_h^{ir}\geq 0$ is the irreversible entropy
production.

4. Adiabatic expansion $D \to A$. Similar to the adiabatic
compression, the working substance is decoupled from the hot
reservoir. During this process, the controlled parameter
$\lambda_b(\tau)$ changes from  $\lambda_b(\tau_c+\tau_a+\tau_h)$ to
$ \lambda_b(\tau_c+\tau_a+\tau_h+\tau_b)$, so the constraint on the
system is loosened.  The entropy production due to the non-adiabatic
dissipation reads
\begin{equation}
\Delta S_b^{ir}  =  S_1 - S_4 \label{ps2}.
\end{equation}

The system recovers to its initial state after a single cycle, and
the total change of entropy of the system is vanishing for a whole
cycle. That is, there exist a following relation:
\begin{equation}
\Delta S  + \Delta S_a^{ir}  + \Delta S_h  + \Delta S_b^{ir}  = 0
\label{ds1},
\end{equation}
where we have defined $\Delta S \equiv \Delta S_c =S_2-S_1$.

Now we follow  the low-dissipation assumption \cite{Esp10} that the
irreversible entropy production during an isothermal process is
assumed to be inversely proportional to the time required for
completing this process, i.e., $\Delta S_\kappa ^{ir} =\Sigma
_\kappa /\tau_\kappa$, where $\Sigma_\kappa$ is a dissipation
constant for the $\kappa$ process with $\kappa = a, b, c ,h$ being
the corresponding thermodynamic processes, respectively. As
emphasized, the irreversible entropy production in any adiabatic
process [$\Delta S_a ^{ir}(\tau_a)$ or $\Delta S_b ^{ir}(\tau_b)$]
cannot be included by the irreversible entropy production in any
isothermal process [$\Delta S_c ^{ir}(\tau_c)$  or $\Delta S_h
^{ir}(\tau_h)$], as lies in the fact that the irreversible entropy
production $\Delta S_\kappa ^{ir}(\tau_\kappa)$ as a function of the
time $\tau_\kappa$ depends on the time taken for the corresponding
process $\kappa$. In contrast to the state variable $\Delta S$ that
depends merely on the initial and final states of the isothermal
processes, here $\Delta S_\kappa ^{ir}$ are process variables
depending on the detailed protocols. As for isothermal processes, we
also adopt the low-dissipation assumption for any adiabatic process
\cite{Fel00, Gor91, pre86, pre85, Rui13} to describe the
irreversible entropy production. It is physically reasonable since
the irreversible entropy production $\Delta S_\kappa^{ir}$ becomes
much smaller and is vanishing in the longtime limit
($\tau_\kappa\rightarrow \infty)$ when the process is quasistatic.

Considering Eqs. (\ref{ts1}), (\ref{ps1}), (\ref{ts2}), (\ref{ps2})
and (\ref{ds1}), the heat $Q_c$ and $Q_h$ are obtained,
\begin{equation}
Q_c  = T_c \left( {\Delta S - \Sigma _c /\tau_c } \right)
\label{qc2},
\end{equation}
and
\begin{equation}
Q_h  = T_h \left( {\Delta S + \Sigma _a /\tau_a  + \Sigma _b /\tau_b
+
 \Sigma _h /\tau_h } \right)
\label{qh2}.
\end{equation}
As a consequence, the work consumed by the system per cycle ($W$)
and the COP  of the refrigeration cycle ($\varepsilon$), are derived
as
\begin{equation}
W=Q_h-Q_c=\left( {T_h  - T_c} \right)\Delta S +
 T_h \Sigma _h /\tau_h+ T_c \Sigma _c /\tau_c+ T_h\left(\Sigma
_a /\tau_a  + \Sigma _b /\tau_b \right), \label{work}
\end{equation}
and
\begin{equation}
\varepsilon= \frac{Q_c}{Q_h-Q_c}=\frac{T_c \left( {\Delta S - \Sigma
_c /\tau_c } \right)}{\left( {T_h  - T_c } \right)\Delta S + T_h
\left( {\Sigma _a /\tau_a  + \Sigma _b /\tau_b +
 \Sigma _h /\tau_h } \right)+ T_c \Sigma _c /\tau_c}
\label{cop}.
\end{equation}
The last term in Eq. (\ref{work}) represents the additional work
 consumed by the system because of the  dissipation in the two
 adiabatic processes. This
additional work to overcome the internally nonadiabatic dissipation
is represented by the two blue areas in Fig. \ref{st}.
\section{Optimum analysis}
In this section we present an optimum analysis of a refrigerator
with internal dissipation which accounts for the irreversible
entropy production during a nonisentropic adiabatic process (The
more details about a nonisentropic adiabatic process can be found in
Ref. \cite{pre86}). If the adiabatic processes are assumed proceed
instantaneously with constant entropy, we recall that \cite{Tom13,
Tu12}: (i) the bounds of the COP under $\Omega$ criterion, between
which there are small differences, are in agreement with the real
experimental data within a range of temperatures of the working
substance; (ii) under the $\chi$ criterion, the upper bound of the
COP fits well with the experimental data, but the COP in the
symmetric limit ($\varepsilon_\chi^{\Sigma_h=\Sigma_c}$) seems to be
considerably larger than the experimental data. In what follows, our
theoretical predictions are expected to agree well with the
experimental data when compared with the experimental data. In
particular, for the $\chi$ criterion, our theoretical data in the
symmetric limit should match more closely with the experimental data
than the ones obtained from the previous models without
consideration of nonadiabatic dissipation \cite{Tom13}.
\subsection{COP at $\chi$ figure of merit}
Substitution of  Eqs. (\ref{work}) and (\ref{cop}) into the $\chi$
figure of merit as the target function, leads to
\begin{equation}
 \chi = \frac{{\varepsilon {{Q_c}} }}{\tau }{ =
 }\frac{{T_c^2 \left( {\Delta S -
 \Sigma _c x_c } \right)^2 }}{{\left[ {\left( {T_h -
 T_c } \right)\Delta S + T_h \left( {\Sigma _a x_a  +
  \Sigma _b x_b  + \Sigma _h x_h } \right){{ +
  }}T_c \Sigma _c x_c } \right]\left( {1/x_a  +
   1/x_b  + 1/x_c  + 1/x_h } \right)}} .\label{chi1}
\end{equation}
Here and hereafter we adopt the variable transformation
 $x_\kappa   = 1/\tau_\kappa  \left( {\kappa  = a, b, c, h}
 \right)$ by taking the inverse of time instead
of the time itself as a variable.

We optimize the  target function $\chi$  over the time variables
$x_\kappa$  to specify the time spent on any thermodynamic process
and also to  maximize this figure of merit. Considering
$\frac{{\partial \chi}}{{\partial x_\kappa }} = 0\left( {\kappa = a,
b, c, h} \right)$, we find the four following relations:
\begin{equation}
\left( {Q_h  - Q_c } \right)x_a x_b x_h  =
 T_c \Sigma _c x_c \left( {2Q_h /Q_c  - 1}
 \right)\left( {x_a x_b x_c  + x_a x_b x_h  +
 x_b x_c x_h  + x_a x_c x_h } \right)
\label{qch1},
\end{equation}
\begin{equation}
\left( {Q_h  - Q_c } \right)x_b x_c x_h  =
 T_h \Sigma _a x_a \left( {x_a x_b x_c  + x_a x_b x_h
  + x_b x_c x_h  + x_a x_c x_h } \right)
\label{qch2},
\end{equation}
\begin{equation}
\left( {Q_h  - Q_c } \right)x_a x_c x_h  =
T_h \Sigma _b x_b \left( {x_a x_b x_c  + x_a x_b x_h  +
 x_b x_c x_h  + x_a x_c x_h } \right)
\label{qch3},
\end{equation}
\begin{equation}
\left( {Q_h  - Q_c } \right)x_a x_b x_c  =
T_h \Sigma _h x_h \left( {x_a x_b x_c  + x_a x_b x_h
+ x_b x_c x_h  + x_a x_c x_h } \right)
\label{qch4}.
\end{equation}
Dividing Eq. (\ref{qch1}) by Eq. (\ref{qch2}), Eq. (\ref{qch3}), Eq.
(\ref{qch4}), respectively, we obtain,
\begin{equation}
\varepsilon^*_\chi  T_h \Sigma _a x_a^2  = \left( {\varepsilon ^
*_\chi
 + 2} \right)T_c \Sigma _c x_c^2
\label{cop1},
\end{equation}
\begin{equation}
\varepsilon^*_\chi   T_h \Sigma _b x_b^2  = \left(
{\varepsilon^*_\chi + 2} \right)T_c \Sigma _c x_c^2 \label{cop2},
\end{equation}
\begin{equation}
\varepsilon^*_\chi  T_h \Sigma _h x_h^2  = \left(
{\varepsilon^*_\chi  + 2} \right)T_c \Sigma _c x_c^2 \label{cop3},
\end{equation}
 Here $\varepsilon^*_\chi $ is the COP under maximum
$\chi$ condition. From Eqs. (\ref{cop1}), (\ref{cop2}), and
(\ref{cop3}), we find that the times spent on the four thermodynamic
processes are optimally distributed as,
\begin{equation}
\tau_\kappa/\tau_h  = \sqrt {\Sigma _\kappa /\Sigma _h }~(\kappa=a,
b), \tau_b/\tau_a = \sqrt {\Sigma _b /\Sigma _a }, \label{tath}
\end{equation}
and
\begin{equation}
\tau_\kappa/\tau_c = \sqrt {T_h \Sigma _\kappa
 /(m T_c \Sigma _c)} ~\left( {\kappa = a,b,h} \right),\label{tatc}
\end{equation}
where $m\equiv\sqrt{{(\varepsilon^*_\chi +2)}/\varepsilon^*_\chi }$
has been adopted and can be determined through numerical calculation
of $\varepsilon_\chi^*$ [see Eqs. (\ref{re1}) and (\ref{alp})
discussed below]. Making summation over Eqs. (\ref{qch1}),
(\ref{qch2}), (\ref{qch3}), and (\ref{qch4}), together with use of
Eqs. (\ref{cop1}), (\ref{cop2}), and (\ref{cop3}), we can derive
after some simple reshuffling,
\begin{equation}
\frac{1}{{\varepsilon^*_\chi }} = \frac{1}{{\varepsilon _C }}
 + \frac{1}{{\varepsilon^*_\chi  +
 \left[ {\left( {2\varepsilon _C  -
 \varepsilon^*_\chi } \right)\alpha\left( {1 + \varepsilon _C } \right)} \right]}}
\label{re1},
\end{equation}
or
\begin{equation}
\varepsilon^*_\chi  = \frac{{\varepsilon _C \left[ {\sqrt {1 +
8\left( {1 + \varepsilon _C } \right)/\alpha}  - 3}
\right]}}{{2\left[ {\left( {1 + \varepsilon _C } \right)/\alpha - 1}
\right]}} \label{re2},
\end{equation}
where  we have used
\begin{equation}
\alpha\equiv \frac{{\Sigma _c x_c }}{{\Sigma _a x_a + \Sigma _b x_b
+ \Sigma _c x_c {\rm{ + }}\Sigma _h x_h }}. \label{alp1}
\end{equation}
It is expected that this result will be reduced to the  one based on
idealized-adiabatic model in which $\Sigma_a=\Sigma_b=0$.  The
expression of $\alpha$ is derived from the more general model in
which
 the nonadiabatic dissipation  and the time spent on any
adiabatic process are involved. Since $0\leq\alpha\leq1$ and
$\varepsilon_C>0$,  $\varepsilon^*_\chi$ increases monotonously as
$\alpha$, and vice versa. As a result, we re-derive the bounds of
the COP at maximum $\chi$ figure of merit \cite{Tu12, Izu13},
\begin{equation}
0 \equiv\varepsilon_\chi^{-}\le \varepsilon ^
*_{\chi}\le\varepsilon_\chi^{+}\equiv \left( {\sqrt {9 + 8\varepsilon _C } -
3} \right)/2 \label{bound1}.
\end{equation}
It is thus clear that the inclusion of the nonadiabatic dissipation
as well as the time taken for the adiabatic process does not change
the upper and lower bounds of the COP at maximum $\chi$ figure of
merit. These lower and upper bounds of $\varepsilon_\chi^*$ are
achieved when $\alpha=0$, and $\alpha=1$, respectively.  Combination
of Eqs. (\ref{cop1}), (\ref{cop2}), (\ref{cop3}), and (\ref{alp1})
can eliminate the ratios $x_\kappa/x_c$ ($\kappa=h, a, b$), leading
to
\begin{equation}
\alpha=\frac{1}{m\left(\sqrt{\frac{T_c\Sigma_h}{T_h\Sigma_c}}
+\sqrt{\frac{T_c\Sigma_a}{T_h\Sigma_c}}+
\sqrt{\frac{T_c\Sigma_b}{T_h\Sigma_c}}\right)+1}, \label{alp}
\end{equation}
with $m=\sqrt{{(\varepsilon^*_\chi +2)}/\varepsilon^*_\chi }$. The
complete asymmetric limits $\Sigma_c/\Sigma_\kappa\rightarrow0$ and
$\Sigma_c/\Sigma_\kappa\rightarrow\infty$,  where $\kappa$
represents $h, a, b$ but except $c$, cause the COP  at maximum
$\chi$ merit of figure to approach  its upper and lower bounds,
$\varepsilon_\chi^{-}=0$, and $\varepsilon_\chi^{+}=\left( {\sqrt {9
+ 8\varepsilon _C }  - 3} \right)/2$, respectively.
\begin{figure}[h]
\includegraphics[width=200pt]{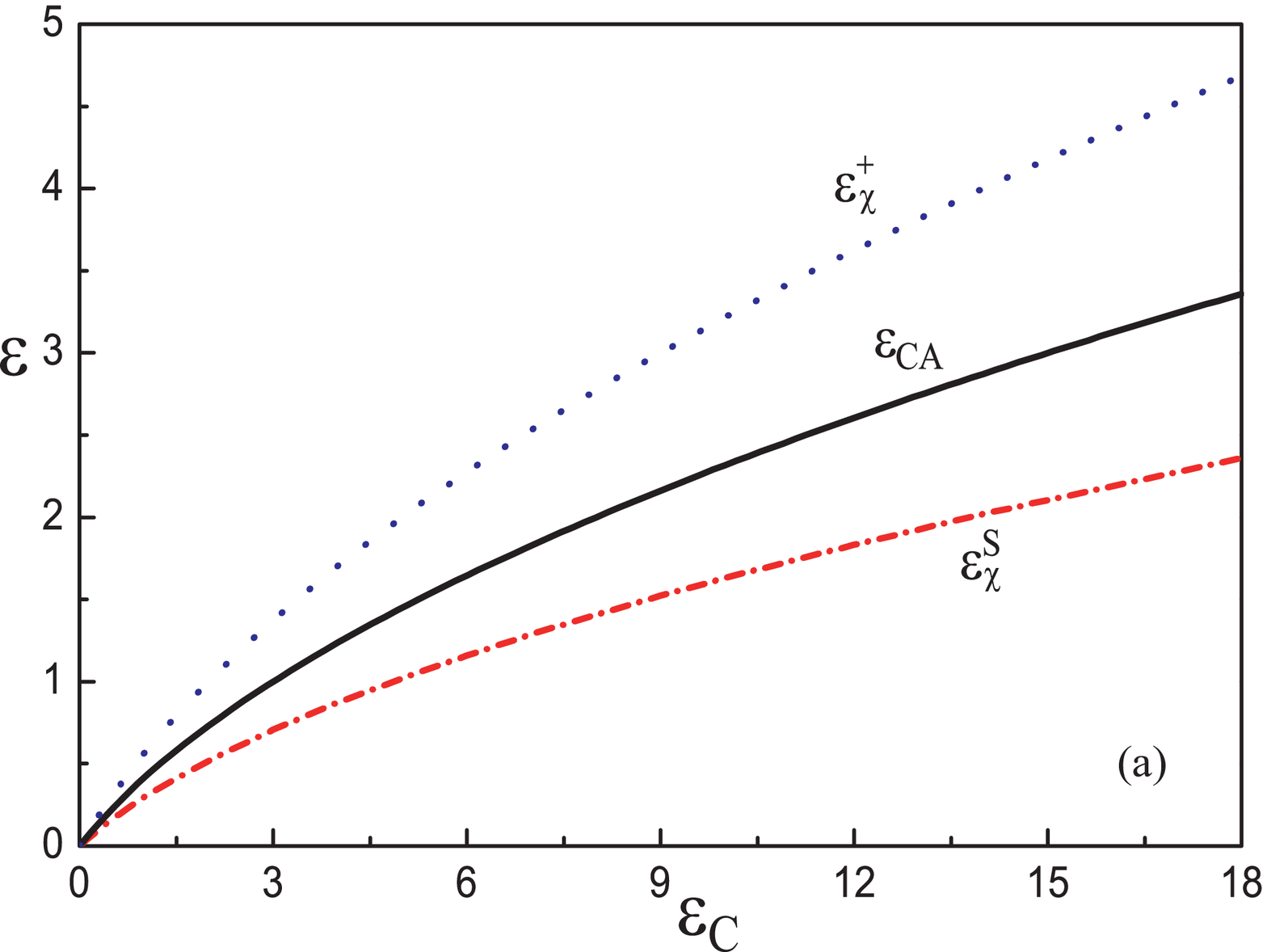}
\includegraphics[width=200pt]{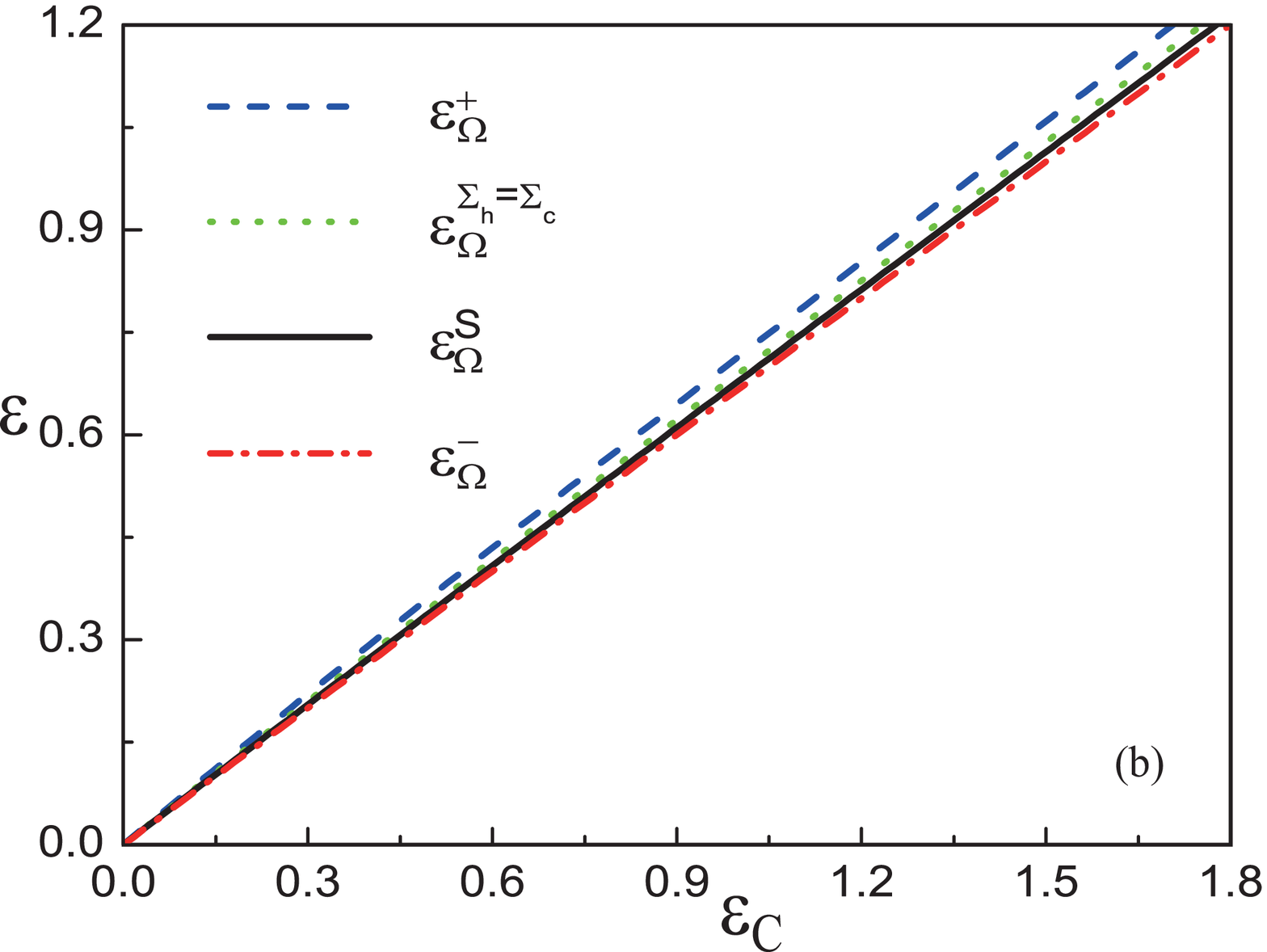}
\caption{(Color online) (a)The values of $\varepsilon_{\chi}$ in the
three limits, $\varepsilon_\chi^+$ (dotted blue line) ,
$\varepsilon_{CA}(=\varepsilon_\chi^{\Sigma_h=\Sigma_c})$ (black
solid line), and $\varepsilon_\chi^{S}$ (red dot-dashed line),
versus the Carnot COP $\varepsilon_{C}$; (b) The values of
$\varepsilon_{\Omega}$ in the four cases, $\varepsilon_\Omega^+$
(blue dashed line), $\varepsilon_\Omega^{\Sigma_h=\Sigma_c}$ (green
dotted line), $\varepsilon_\Omega^{S}$ (black solid line), and
$\varepsilon_\Omega^{-}$ (red dot-dashed line), versus the Carnot
COP $\varepsilon_{C}.$} \label{cm}
\end{figure}

 When the dissipations in the  two adiabatic and two isothermal processes are
symmetric, respectively, we have
$\Sigma_a=\Sigma_b=r\Sigma_h=r\Sigma_c$, with $r$ being the ratio.
In such a case we consider three special situations: (i)
\emph{$r\rightarrow0$}. The  nonadiabatic dissipations for the two
adiabatic processes are vanishing, while the dissipations during the
two isothermal processes are symmetric. By using Eq. (\ref{re2}),
the CA COP is recovered,
$\varepsilon_\chi^{\Sigma_h=\Sigma_c}=\varepsilon_{CA}= \sqrt {1 +
\varepsilon _C }
 - 1 $, which is also the upper bound of the COP in such a case. (ii)
\emph{ $r\rightarrow\infty$.} The lower bound of the COP is
achieved,
 $\varepsilon_\chi^{-}=0$. (iii)\emph{ $r=1$}. The dissipations in four
thermodynamic processes are symmetric. Here
$\varepsilon_\chi^S\equiv\varepsilon_\chi(r=1)$ is defined for
convenience, and its value can be done numerically based on Eqs.
(\ref{re1}) and (\ref{alp}) for any given value of $T_h/T_c$ (i.e.,
the value of $\varepsilon_C$). At the super symmetric limit we
obtain readily from Eqs. (\ref{tath}) and (\ref{tatc}) that the time
ratios of $\tau_\kappa/\tau_c$ $(\kappa = a,~b,~h)$ are
$\tau_\kappa/\tau_c = \sqrt {T_h/(m T_c)}$ with
$m=\sqrt{{(\varepsilon^*_\chi +2)}/\varepsilon^*_\chi }$, and that
the time allocations to the rest three processes are
equal($\tau_h=\tau_a=\tau_b)$.
 In Fig. \ref{cm} (a) we plot the COP
$\varepsilon_\chi^S$ as a function of $\varepsilon_{C}$, comparing
$\varepsilon_{CA}$ with the upper
bound $\varepsilon_\chi^{+}$ of the Carnot-like refrigeration cycle. %Fig shows that, the values of COP at
%maximum $\chi$ figure of merit, match the ones of real refrigerators
%more closely than the ones derived from the previous idealized model
%under the assumption that the non-adiabatic dissipations on the
%adiabatic process were negligible.

\subsection{COP at maximum ${\dot{\Omega}}$ figure of merit}
The $\Omega$ criterion, a trade-off between maximum cooling and lost
cooling loads, is defined as $\Omega=\left( {2\varepsilon  -
\varepsilon _{\max } } \right){W}$ \cite{Her01}. The target
function, $\dot{\Omega }
 = \left( {2\varepsilon  - \varepsilon _{\max } }
 \right)\frac{W}{\tau }$, can be expressed
 as
 \begin{equation}
\dot{\Omega}
 =\left[ {2Q_c  - \varepsilon _C
 \left( {Q_h  - Q_c } \right)} \right]\left( {x_a x_b x_c
 + x_a x_b x_h  + x_b x_c x_h  +
 x_a x_c x_h } \right)/x_a x_b x_c x_h,
 \label{tf2}
\end{equation}
where we have made the variable transformation
$x_\kappa=1/\tau_\kappa$ ($\kappa=h, c, a, b$).
%For refrigerators, the $\Omega $ criterion, a trade-off between maximum
% cooling load and minimum lost cooling load. As for refrigerators,
% it's completely a new word put forward by C.de Tomas \cite{cdtom},
% who use his experiment dates and summary it subtly. Now we use
%the same way to study a more universal result with having taken
%consideration of the time and irreversibility entropy in the
%other two adiabatic processes.
%, with $\tau=\tau_h+\tau_c+\tau_a+\tau_b$ the cycle time,
% can be written as by the use of variable transformation
Setting the derivatives of $\dot{\Omega}$ with respect to $x_\kappa$
($\kappa = h,c,a,b$) equal to zero, we derive the optimal equations:
\begin{equation}
\left[ {2Q_c  - \varepsilon _C \left( {Q_h  -
Q_c } \right)} \right]\frac{{x_b x_c x_h }}{{x_a }}
 = T_h \Sigma _a \varepsilon _C \left( {x_a x_b x_c
  + x_a x_b x_h  + x_b x_c x_h  + x_a x_c x_h } \right)
\label{qx1},
\end{equation}
\begin{equation}
\left[ {2Q_c  - \varepsilon _C \left( {Q_h  -
Q_c } \right)} \right]\frac{{x_a x_c x_h }}{{x_b }}
= T_h \Sigma _b \varepsilon _C \left( {x_a x_b x_c
+ x_a x_b x_h  + x_b x_c x_h  + x_a x_c x_h } \right)
\label{qx2},
\end{equation}
\begin{equation}
\left[ {2Q_c  - \varepsilon _C \left( {Q_h  -
Q_c } \right)} \right]\frac{{x_a x_b x_c }}{{x_h }}
= T_h \Sigma _h \varepsilon _C \left( {x_a x_b x_c
 + x_a x_b x_h  + x_b x_c x_h  + x_a x_c x_h } \right)
\label{qx3},
\end{equation}
\begin{equation}
\left[ {2Q_c  - \varepsilon _C \left( {Q_h  -
Q_c } \right)} \right]\frac{{x_a x_b x_h }}{{x_c }}
= T_c \Sigma _c \left( {2 + \varepsilon _C
} \right)\left( {x_a x_b x_c  + x_a x_b x_h  +
x_b x_c x_h  + x_a x_c x_h } \right)
\label{qx4}.
\end{equation}
 Dividing Eq. (\ref{qx4}) by
Eqs. (\ref{qx1}), (\ref{qx2}), and (\ref{qx3}), respectively, we
have
\begin{equation}
\frac{{x_c }}{{x_a }} = \sqrt {\frac{{\Sigma _a
\left( {1 + \varepsilon _C } \right)}}{{\Sigma _c
 \left( {2 + \varepsilon _C } \right)}}}
\label{bx1},
\end{equation}
\begin{equation}
\frac{{x_c }}{{x_b }} = \sqrt {\frac{{\Sigma _b
\left( {1 + \varepsilon _C } \right)}}{{\Sigma _c
\left( {2 + \varepsilon _C } \right)}}}
\label{bx2},
\end{equation}
\begin{equation}
\frac{{x_c }}{{x_h }} = \sqrt {\frac{{\Sigma _h
\left( {1 + \varepsilon _C } \right)}}{{\Sigma _c
\left( {2 + \varepsilon _C } \right)}}}
\label{bx3}.
\end{equation}
It follows, substitution of $\tau_\kappa=1/x_\kappa~ (\kappa=h, c, a
, b)$ into Eqs. (\ref{bx1}), (\ref{bx2}),  and (\ref{bx3}), that the
optimal ratios of the time $\tau_\kappa/\tau_h~(\kappa=a, b)$ as
well as $\tau_b/\tau_a$ are still given by Eq. (\ref{tath}), but
that under the $\Omega$ criterion the time ratio
$\tau_\kappa/\tau_c~(\kappa=h,a, b)$ becomes
\begin{equation}
\frac{\tau_\kappa}{\tau_c}  = \sqrt {\frac{\Sigma _\kappa \left( {1
+ \varepsilon _C } \right)}{\Sigma _h \left( {2 + \varepsilon _C }
\right)}} ,\left( {\kappa = a,b,h} \right) .\label{tautc}
\end{equation}

  Directly adding both sides of Eqs. (\ref{qx1}), (\ref{qx2})
(\ref{qx3}), and (\ref{qx4}), and using Eqs. (\ref{bx1}),
(\ref{bx2}),  and (\ref{bx3}), leads to the result as
\begin{equation}
\Delta S = 2\left( {2 + \varepsilon _C }
\right)\Sigma _c x_c  + 2\left( {1 + \varepsilon _C }
\right)\left( {\Sigma _a x_a  + \Sigma _b x_b  +
 \Sigma _h x_h } \right)
\label{deltas1}.
\end{equation}
 It follows, substituting Eq. (\ref{deltas1}) into  Eq. (\ref{cop}), that the COP
at maximum $\dot{\Omega}$ conditions is
%\begin{equation}
%\varepsilon  = \frac{{Q_c }}{W} = \frac{{Q_c }}{{Q_h  -
% Q_c }} = \frac{{T_c \left( {\Delta S -
%\Sigma _x x_c } \right)}}{{\left( {T_h  -
% T_c } \right)\Delta S + T_c \Sigma _c x_c
% + T_h \left( {\Sigma _a x_a  + \Sigma _b x_b
%+ \Sigma _h x_h } \right)}}
%\label{result3}.
%\end{equation}
%and taking into account Eqs. (\ref{bx1}), (\ref{bx2}), (\ref{bx3}),
%we can obtain the COP under maximum  in terms of the temperatures
%and dissipations constants $\Sigma _\kappa  \left( {\kappa  = a, b,
%c, h} \right)$, read as
\begin{equation}
\varepsilon _\Omega^*   = \frac{{3 + 2\varepsilon _C  + 2\gamma}}{{4
+
 3\varepsilon _C  + 3\gamma}}\varepsilon _C
\label{result4},
\end{equation}
where $\gamma = \sqrt {\left( {1 + \varepsilon _C } \right)\left( {2
+ \varepsilon _C } \right)\Sigma _a /\Sigma _c }
  + \sqrt {\left( {1 + \varepsilon _C }
  \right)\left( {2 + \varepsilon _C }
  \right)\Sigma _b /\Sigma _c }  +
  \sqrt {\left( {1 + \varepsilon _C }
  \right)\left( {2 + \varepsilon _C } \right)\Sigma _h /\Sigma _c }
  $, which simplifies to $\gamma=\sqrt {\left( {1 + \varepsilon _C }
  \right)\left( {2 + \varepsilon _C } \right)\Sigma _h /\Sigma _c }$
  in the ideal-adiabatic refrigeration cycle. The value of $\gamma$
is a non-negative number, varying from $0$ to $\infty$.
  Hence, the COP at maximum $\dot{\Omega}$ figure of merit,
  $\varepsilon_{\Omega}$, must be situated between
\begin{equation}
\varepsilon _\Omega ^ -   \equiv \frac{2}{3} \varepsilon _C  \le
\varepsilon _\Omega^*   \le
 \frac{{3 + 2\varepsilon _C }}{{4 + 3\varepsilon _C }}
 \varepsilon _C  \equiv \varepsilon _\Omega ^ +
\label{bound2}.
\end{equation}
The upper and lower bounds for the optimized COP at maximum
$\dot{\Omega}$ figure of merit, $\varepsilon _\Omega ^ -$ and
$\varepsilon _\Omega ^ +$, versus the Carnot COP $\varepsilon_C$,
are plotted in Fig. \ref{cm}(b).
\begin{figure}[h]
\includegraphics[width=250pt]{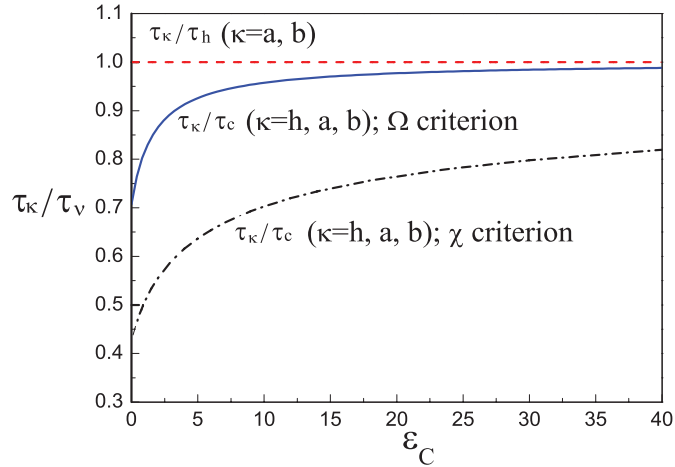}
\caption{(Color online) The ratios of $\tau_\kappa/\tau_\nu$
($\kappa=h, a, b$, and $\nu=h, c$) within maximum $\chi$ as well as
$\dot{\Omega}$ figure of merit versus the Carnot COP $\varepsilon_C$
at the super symmetric limit.  Here the optimal values of
$\tau_\kappa/\tau_c$ ($\kappa=h, a, b$) under
 $\chi$ and $\Omega$ criteria, are indicated by a back dash-dotted
a blue solid  line, respectively.  The values of
$\tau_\kappa/\tau_h$ ($\kappa=a, b$) are equal to $1$ both under
$\chi$ and $\Omega$ criteria, and are represented by a red dashed
line. } \label{tr}
\end{figure}

As in the case of the $\chi$ figure of merit,
 the expression of COP at maximum $\dot{\Omega}$ condition similar to the
  corresponding one obtained in the model \cite{Tom13} with idealized adiabatic
  processes, and the internally nonadiabatic dissipation has no influence on the bounds of the COP.
Here the optimal value of COP, however,  represents  a broader
context by including the nonadiabatic dissipation and the time
required for completing any adiabat.  When adiabatic processes
proceeds simultaneously and are isentropic ($\Sigma_a=\Sigma_b=0$),
our result is reduced to that in \cite{Tom13}, as expected.

 If the dissipations of the two adiabatic and two isothermal
processes are symmetric, respectively, i.e., $\Sigma _a = \Sigma _b
= r\Sigma _c = r\Sigma _h $, then $\gamma = (2\sqrt r
 + 1)\sqrt {\left( {1 + \varepsilon _C } \right)
 \left( {2 + \varepsilon _C } \right)} $, and
Eq. (\ref{result4}) becomes
\begin{equation}
 \varepsilon _\Omega (r)
  = \frac{{3 + 2\varepsilon _C  +
\left( {4\sqrt r  + 2} \right) \sqrt {\left( {1 + \varepsilon _C }
\right)\left( {2 + \varepsilon _C } \right)} }}{{4 + 3\varepsilon _C
+ \left( {6\sqrt r   + 3} \right) \sqrt {\left( {1 + \varepsilon _C
} \right)\left( {2 + \varepsilon _C } \right)} }}\varepsilon _C
\label{symr}.
\end{equation}
From Eq. (\ref{symr}), we find in such a case that the bounds of the
COP at maximum $\dot{\Omega}$ figure of merit are achieved,
$\frac{2}{3} \varepsilon _C  \le \varepsilon _\Omega(r) \le\frac{{3
+ 2\varepsilon _C
   + 2\sqrt {\left( {1 + \varepsilon _C } \right)
  \left( {2 + \varepsilon _C } \right)} }}{{4 +
   3\varepsilon _C  + 3\sqrt {\left( {1 +
    \varepsilon _C } \right)\left( {2 +
    \varepsilon _C } \right)} }}\varepsilon _C$, when
    $r\rightarrow0$ and $r\rightarrow \infty$, respectively.
In  the particular case when the dissipations of the four
thermodynamic processes are symmetric, the COP can be obtained by
the use of $r=1$,
\begin{equation}
 \varepsilon ^{S}_\Omega
 = \frac{{3 + 2\varepsilon _C  + 6\sqrt {
 \left( {1 + \varepsilon _C } \right)\left(
 {2 + \varepsilon _C } \right)} }}{{4 +
3\varepsilon _C  + 9\sqrt {\left( {1 + \varepsilon _C }
\right)\left( {2 + \varepsilon _C } \right)} }}\varepsilon _C.
\label{vanc}
\end{equation}
Then  the optimal time ratios of $\tau_\kappa/\tau_c~ (\kappa=h, a,
b)$ in Eq. (\ref{tautc}) simplifies to $\tau_\kappa/\tau_c
=\sqrt{T_c/(2T_h-T_c)}~(\kappa=h, a, b) $ in this super asymmetric
case, while the optimized times spent on the other three processes
are equal ($\tau_a=\tau_b=\tau_h)$.  At the super symmetric limit,
the time ratios of $\tau_\kappa/\tau_\nu$, with $\kappa=h, a, b$ and
$\nu=h,c$ as functions of the Carnot COP $\varepsilon_C$, under
$\Omega$ and $\chi$ criteria, are plotted in Fig. \ref{tr} by using
Eqs. (\ref{tath}), (\ref{tatc}), and (\ref{tautc}).   Fig. \ref{tr}
shows that, whether under $\chi$ or $\Omega$ criterion, the time
taken for the cold isothermal process is larger than the ones for
the other three processes, on which the times spent are equal to
each other. This result is contrast to the fact that, for an
irreversible heat engine \cite{pre86}, the hot isothermal process
proceeds most slowly during a cycle, with equal times required for
completing the cold isothermal and two adiabatic processes. This is
not surprising, since the heat is transported into the system during
the cold (hot) isothermal process for the refrigerator (heat
engine), and the additional heat developed by the nonadiabatic
dissipation is related to the high temperature $T_h$ (low
temperature $T_c$) for the refrigerator (heat engine). For the model
with idealized adiabatic processes ($\Sigma_a=\Sigma_b=0$), the
symmetric limit ($r=1$) gives rise to the simple form of Eq.
(\ref{result4}),
\begin{equation}
\varepsilon_\Omega^{\Sigma_c=\Sigma_h}\equiv\varepsilon_\Omega
^{\Sigma _h  = \Sigma _c }(\Sigma_a=\Sigma_b=0) = \frac{{\varepsilon
_C }}{{\sqrt {\left( {1 + \varepsilon _C } \right)\left( {2 +
\varepsilon _C } \right)} - \varepsilon _C}} \label{sym3}.
\end{equation}
At the symmetric limits (with and without nonadiabatic dissipation)
the optimal COP's, $\varepsilon_\Omega^{S}$ determined according to
by Eqs. (\ref{vanc}) and $\varepsilon_\Omega^{\Sigma_c=\Sigma_h}$
given by Eq. (\ref{sym3}), are also shown in Fig. \ref{cm} (b). It
is clear from Fig. \ref{cm} (b) that the nonadiabatic dissipation
leads to a very slight decrease in the COP.

\subsection{Comparison between our prediction with experimental
data}

\begin{figure}[h]
\includegraphics[width=250pt]{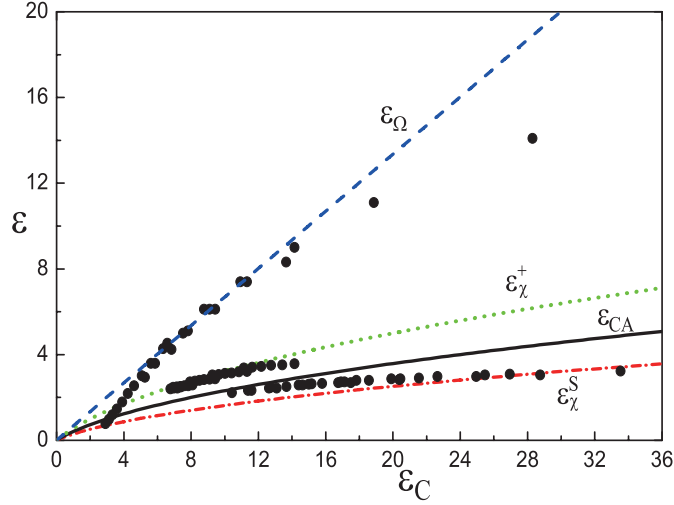}
\caption{(Color online) Comparison between theoretical results
(lines) and three sets of experimental results (points).  Here
$\varepsilon_\Omega$, $\varepsilon_\chi^+$, $\varepsilon_{CA}$, and
$\varepsilon_\chi^S$, are indicated by a blue dashed line, a green
dotted line,a black solid line, and a red dot-dashed line,
respectively.  The values in the four cases of
$\varepsilon_{\Omega}$, $\varepsilon_\Omega^+$,
$\varepsilon_\Omega^{\Sigma_h=\Sigma_c}$, $\varepsilon_\Omega^{S}$,
and $\varepsilon_\Omega^{-}$, are indistinguishable and collapse
into a single curve in this plotted scale.
 } \label{te}
\end{figure}

It would be instructive to compare our theoretical predictions with
the observed COP's of  some real refrigerators.   Our theoretical
prediction versus the data of the real refrigerators \cite{Gor00} at
different values of temperature are plotted in Fig. \ref{te}, which
shows that the theoretical results agree well with the experimental
refrigerator data, whether at maximum $\chi$ or $\dot{\Omega}$
figure of merit. Applying the $\Omega$ criterion to optimization on
the refrigerator cycle, we find that there are relatively small
differences even between the lower and upper bounds
($\varepsilon_\Omega^+$ and $\varepsilon_\Omega^-$) of the COP for
the refrigerator cycle. The values of COP, $\varepsilon_\Omega^+,
 \varepsilon_\Omega^{\Sigma_h=\Sigma_c}$, $\varepsilon_\Omega^S$, and $\varepsilon_\Omega^-$
 are indistinguishable in the plotted scale of Fig. \ref{te} and
 are
in good agreement with experimental data, particulary for some
values of $\varepsilon_C$. Under maximum $\chi$ condition, our
calculation of COP under maximum $\chi$ in the symmetric limit,
$\varepsilon_\chi^S$, match more closely with the experimental data
than the corresponding ones obtained in the previous model with
idealized adiabatic processes,
$\varepsilon_{CA}$=($\varepsilon_\chi^{\Sigma_h=\Sigma_c}$), as
expected. Hence, our result suggests that internally nonadiabatic
dissipation indeed induces the effects on the performance in the
heat devices and thus can not be negligible in comparison with the
experimental data.
\section{conclusion}
 In conclusion, we have
analyzed the COP at $\chi$ and $\dot{\Omega}$ figure of merits for
an irreversible Carnot-like refrigerator with non-adiabatic
dissipation.  In the limits of extremely asymmetric dissipations,
the COP either at maximum $\chi$ or at $\Omega$ figure of merit,
converges to the same bounds as the corresponding ones obtained from
previous models with idealized adiabatic processes. When the
dissipations in two isothermal and two adiabatic processes are
symmetric, respectively, comparison between our theoretical
predictions of COP at maximum $\chi$ figure of merit and the
observed COP's of real refrigerators shows that our values matches
more closely than the ones derived in previous models with no
inclusion of non-adiabatic dissipation.

\emph{Acknowledgements:} We gratefully acknowledge the financial
support from the National Natural Science Foundation of China under
Grant No. 11265010, No. 11065008, and No. 11191240252; the State Key
Programs of China under Grant No. 2012CB921604; the Jiangxi
Provincial Natural Science Foundation under Grant No.
20132BAB212009, China; and MICIN (Spain) under Grant No.
FIS2010-17147FEDER. We are very grateful to Prof. Zhanchun Tu at
Beijing Normal University for his valuable comments on the
manuscript.

\end{document}